\documentclass[prc, reprint, amsmath, amssymb, superscriptaddress, nofootinbib, showpacs]{revtex4-1}

\usepackage{graphicx}%
\usepackage{multirow}
\usepackage{float}
\usepackage{epstopdf}
\usepackage[utf8]{inputenc}
\usepackage[english]{babel}
\usepackage{footnote}

\begin{document}

\title{Identification of a potential ultra-low Q value electron capture decay branch in $^{75}$Se via a precise Penning trap measurement of the mass of $^{75}$As}

\author{M. Horana Gamage}
\email{corresponding author: redsh1m@cmich.edu}
\affiliation{Department of Physics, Central Michigan University, Mount Pleasant, Michigan, 48859, USA}

\author{R. Bhandari}
\affiliation{Department of Physics, Central Michigan University, Mount Pleasant, Michigan, 48859, USA}

\author{G. Bollen}
\affiliation{Facility for Rare Isotope Beams, East Lansing, Michigan, 48824, USA}
\affiliation{Department of Physics and Astronomy, Michigan State University, East Lansing, Michigan 48824, USA}

\author{N. D. Gamage}
\affiliation{Facility for Rare Isotope Beams, East Lansing, Michigan, 48824, USA}

\author{A. Hamaker}
\affiliation{National Superconducting Cyclotron Laboratory, East Lansing, Michigan, 48824, USA}
\affiliation{Department of Physics and Astronomy, Michigan State University, East Lansing, Michigan 48824, USA}

\author{D. Puentes}
\affiliation{National Superconducting Cyclotron Laboratory, East Lansing, Michigan, 48824, USA}
\affiliation{Department of Physics and Astronomy, Michigan State University, East Lansing, Michigan 48824, USA}

\author{M. Redshaw}
\affiliation{Department of Physics, Central Michigan University, Mount Pleasant, Michigan, 48859, USA}
\affiliation{National Superconducting Cyclotron Laboratory, East Lansing, Michigan, 48824, USA}

\author{R. Ringle}
\affiliation{Facility for Rare Isotope Beams, East Lansing, Michigan, 48824, USA}

\author{S. Schwarz}
\affiliation{Facility for Rare Isotope Beams, East Lansing, Michigan, 48824, USA}

\author{C. S. Sumithrarachchi}
\affiliation{Facility for Rare Isotope Beams, East Lansing, Michigan, 48824, USA}

\author{I. Yandow}
\affiliation{National Superconducting Cyclotron Laboratory, East Lansing, Michigan, 48824, USA}
\affiliation{Department of Physics and Astronomy, Michigan State University, East Lansing, Michigan 48824, USA}
\date{\today}%

\begin{abstract}
\begin{description}
\item[Background]
Low energy $\beta$ and electron capture (EC) decays are important systems in neutrino mass determination experiments. An isotope with an ultra-low Q value $\beta$-decay to an excited state in the daughter with $Q_{es}$ $<$ 1 keV could provide a promising alternative candidate for future experiments. $^{75}$Se EC and $^{75}$Ge $\beta$-decay represent such candidates, but a more precise determination of the mass of the common daughter, $^{75}$As, is required to evaluate whether their potential decay branches are energetically allowed and ultra-low.
\item[Purpose]
Perform a precise atomic mass measurement of $^{75}$As and combine the result with the precisely known atomic masses of $^{75}$Se and $^{75}$Ge, along with nuclear energy level data for $^{75}$As to evaluate potential ultra-low Q value decay branches in the EC decay of $^{75}$Se and the $\beta$-decay of $^{75}$Ge.
\item[Method]
The LEBIT Penning trap mass spectrometer at the Facility for Rare Isotope Beams was used to perform a high-precision measurement of the atomic mass of $^{75}$As via cyclotron frequency ratio measurements of $^{75}$As$^{+}$ to a $^{12}$C$_{6}^{+}$ reference ion. 
\item[Results]
The $^{75}$As mass excess was determined to be ME($^{75}$As) = --73 035.98(43) keV, from which the ground-state to ground-state Q values for $^{75}$Se EC and $^{75}$Ge $\beta$-decay were determined to be 866.50(44) keV and 1179.01(44) keV, respectively. These results were compared to energies of excited states in $^{75}$As at 865.4(5) keV and 1172.0(6) keV to determine Q values of 1.1(7) keV and 7.0(7) keV for the potential ultra-low EC and $\beta$-decay branches of $^{75}$Se and $^{75}$Ge, respectively.
\item[Conclusion]
The candidate ultra-low Q value EC and $\beta$-decay branches of $^{75}$Se and $^{75}$Ge to excited states in $^{75}$As, identified by Gamage, $\textit{et al.}$ [N.D. Gamage, Hyperfine Interact. $\textbf{240}$, 43 (2019)] are seen to be energetically allowed. The $^{75}$Se EC decay to the 865.4 keV excited state in $^{75}$As is potentially ultra-low with $Q_{es}$ $\approx$ 1 keV. However, a more precise determination of the 865.4(5) keV level in $^{75}$As is required to draw a more definite conclusion about the energy of this potential decay channel.
\end{description}
\end{abstract}

\maketitle

\section{Introduction}

The ongoing development of sensitive, high-precision experimental techniques and sophisticated theoretical calculations has continued to make nuclear $\beta$-decay an important tool in modern physics, for example, in tests of the Standard Model (SM) and searches for physics beyond the SM~\cite{Har2015,Sev2006,Bur2022}, and in investigations of the nature and absolute mass of the neutrino~\cite{Avi2008,Fri2021,Aker2019,Monreal2009_Project8,Gastaldo2017,Faverzani2016}.

Extreme cases of $\beta$-decay e.g. highly-forbidden decays and very low energy decays, pose significant experimental challenges in the detection of these rare processes, and theoretical challenges in describing them. Ultra-low (UL) Q value $\beta$-decays, in which the parent nucleus decays to an excited state in the daughter with $Q_{es}$ $<$ 1 keV, provide a particularly interesting case, since they have both experimental and theoretical implications. An isotope with an ultra-low Q value $\beta$-decay could potentially be used as a new candidate in direct neutrino mass determination experiments~\cite{cat2005,cat2007,Suh2014,kop2010}. They can also provide a powerful tool to test the role of atomic interference effects in nuclear $\beta$-decay~\cite{mus2010,Suh2010}, which become more significant for very low energy decays.

To date, only one ultra-low Q value $\beta$-decay has been observed, that of $^{115}$In to the 3/2$^{+}$ first excited state in $^{115}$Sn~\cite{cat2005}. The confirmation that this decay branch is energetically allowed was made by precise Penning trap measurements of the $^{115}$In -- $^{115}$Sn mass difference at Florida State University~\cite{mou2009} and with JYLFTRAP at the University of Jyv{\"a}skyl{\"a}~\cite{wie2009}, combined with the precisely known energy of the 3/2$^{+}$ level in $^{115}$Sn~\cite{Bla2005,Zhe2019}. Using the results of Ref.~\cite{mou2009} and \cite{Zhe2019}, the resulting $Q_{\textrm{UL}}$ = 0.147(10) keV, making this the lowest known Q value $\beta$-decay. Theoretical calculations of the partial half-life for $^{115}$In decay to $^{115}$Sn(3/2$^{+}$) showed a significant discrepancy compared to experimental results~\cite{mus2010,Suh2010}. To identify potential candidates for future direct neutrino mass determination experiments, and to further theoretical developments, additional ultra-low Q value decays are required to be identified and to have their Q values and partial half-lives measured.

A large number of potential ultra-low Q value decays do in fact exist---see for example Refs.~\cite{mus2010_2,Suh2014,kop2010,Haa2013,Mus2011}, and  Ref.~\cite{gam2019} for a recent comprehensive evaluation. The identification of these ultra-low Q value decay candidates is made from a comparison of the ground-state to ground-state (gs-gs) Q value, $Q_{gs}$, obtained from atomic mass data for parent and daughter isotopes~\cite{Wang2021}, and nuclear energy level data for the daughter~\cite{nndc}. In order to determine if a decay to an excited state in the daughter is energetically allowed and ultra-low, both $Q_{gs}$, and the energy of the final state in the daughter, $E^{*}$, must be known to $\ll$ 1 keV. While nuclear energy levels are often known to precisions of $\sim$10 -- 100 eV, atomic masses are usually not, except in cases where they have been measured precisely by dedicated Penning trap experiments.

In the last few years, there has been growing interest in performing precise Penning trap measurements to determine $Q_{es}$ values for potential ultra-low Q value decay candidates that have been identified. Studies have been performed on $^{131}$Cs~\cite{Kar2019_131Cs} at ISOLTRAP; $^{135}$Cs~\cite{deRoubin2020}, $^{72}$As~\cite{Ge2021}, and $^{159}$Dy~\cite{Ge2021_159Dy} at JYFLTRAP; $^{112,113}$Ag and $^{115}$Cd with the Canadian Penning Trap~\cite{Gam2022_CPT}, and $^{89}$Sr and $^{139}$Ba at LEBIT~\cite{Sandler2019_89Sr}. These measurements showed that $^{135}$Cs could have an ultra-low Q value $\beta$-decay branch, and that in $^{159}$Dy EC an experimentally confirmed~\cite{Myslek1969} decay to the 11/2$^{+}$ state in the daughter has a $Q_{es}$ = 1.2(2) keV. The other candidates were ruled out.

In this paper we report on precise determinations of the $Q_{gs}$ values for $^{75}$Ge $\beta$-decay and $^{75}$Se EC decay via a precise measurement of the mass of the common daughter, $^{75}$As. These candidates were first identified in Ref.~\cite{gam2019}, along with $^{89}$Sr and $^{139}$Ba~\cite{Sandler2019_89Sr}, as cases where the mass of the stable daughter isotope was known less precisely than that of the unstable parents, making them ideal cases for study with the offline ion sources available at Penning trap facilities such as LEBIT. 

\begin{figure}[h]
    \includegraphics[width=1\columnwidth]{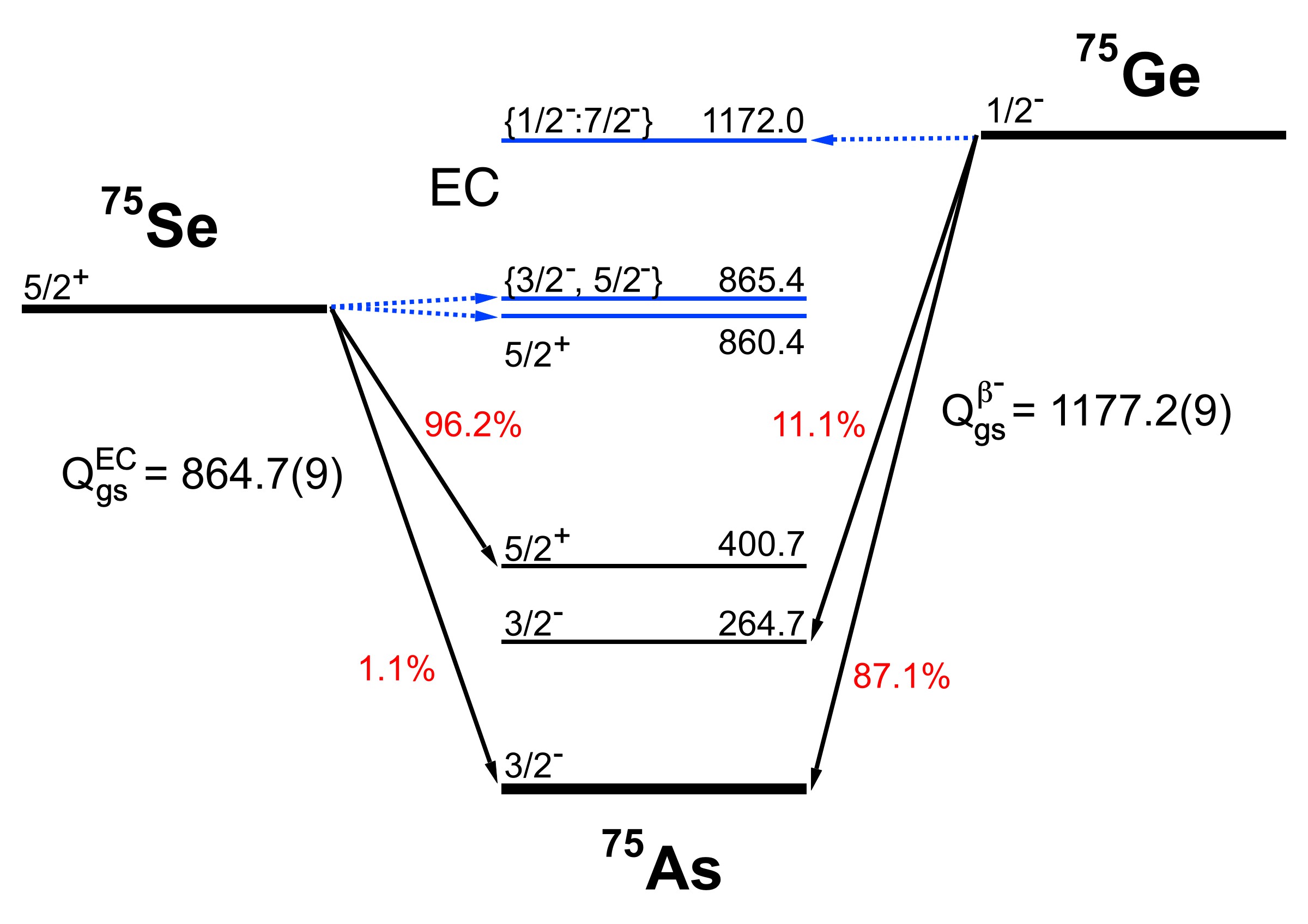}
    \caption{(color online) Decay scheme for $^{75}$Se EC and $^{75}$Ge $\beta$-decay. The main decay branches are indicated by solid black arrows with corresponding branching ratios. The potential ultra-low Q value decay branches are indicated by dashed blue arrows. Spin assignments in braces indicate the range of possible values for that level. The ground-state to ground-state Q values are obtained using data from the AME2020~\cite{Wang2021}. Energy levels and Q values are given in units of keV.}
    \label{fig:levels}
\end{figure}

Fig.~\ref{fig:levels} shows the energy levels and decay schemes for the $A = 75$ Ge-As-Se triplet. The potential ultra-low Q value decays are indicated by dashed blue arrows. For $^{75}$Ge, the potential ultra-low Q value $\beta$-decay is to the 1172.0(6) keV level in $^{75}$As, identified as a negative parity state with spin between 1/2 and 7/2~\cite{nndc}. For $^{75}$Se, there are two potential ultra-low Q value EC decay branches: to the 865.4(5) keV level, identified as having $J^{\pi}$ = 3/2$^{-}$ or 5/2$^{-}$, and to the 860.4(4) keV 1/2$^{+}$ level. The $Q_{gs}$ values listed in Fig.~\ref{fig:levels} are calculated using data from the most recent atomic mass evaluation, AME2020~\cite{Wang2021}, and are limited by the 0.88 keV/$c^{2}$ uncertainty in the mass of $^{75}$As. The uncertainties in the masses of $^{75}$Ge and $^{75}$Se are 0.052 keV/$c^{2}$ and 0.073 keV/$c^{2}$, respectively, and contribute negligibly to the uncertainty in the Q value. The resulting potential ultra-low Q value decay energies to the respective excited states in $^{75}$As are $Q^{\beta}_{es}$($^{75}$Ge) = 5.2(1.1) keV and $Q^{\textrm{EC}}_{es}$($^{75}$Se) = --0.7(1.0) keV and 4.3(1.0) keV. Hence, a new Penning trap measurement of the $^{75}$As mass is called for to check the accuracy of the Q values and eliminate uncertainty due to current uncertainty in the $^{75}$As mass.

\section{Experimental Description}\label{Section:Exp}
The Low Energy Beam and Ion Trap (LEBIT)~\cite{Ringle2009,rin2013}, located at the Facility for Rare Isotope Beams (FRIB) was used to perform precise cyclotron frequency ratio measurements of singly-charged $^{75}$As$^{+}$ and $^{12}$C$_{6}^{+}$ carbon cluster ions with a similar mass-to-charge ratio. The $^{75}$As$^{+}$ ions were delivered to the LEBIT facility from a batch mode ion source (BMIS) where they were produced as a by-product during normal BMIS operation. The $^{12}$C$_{6}^{+}$ ions were produced with a laser ablation source (LAS)~\cite{izz2016} in which a Sigradur\textsuperscript{\textregistered}~\cite{sigradur} glassy carbon target was installed. In either case, ions first enter the LEBIT cooler and buncher~\cite{Sch2016}, in which they are thermalized by a helium buffer gas, confined in a linear rf Paul trap, and then released as low emittance, $\sim$100 ns length pulses. Ions are then directed toward the Penning trap, which has a hyperbolic structure and is housed in a 9.4 T magnetic field. 

In the Penning trap, ions undergo three normal modes of motion: reduced-cylotron, magnetron, and axial, with frequencies $f_{+}$, $f_{-}$, and $f_{z}$, respectively. The three normal mode frequencies can be combined to determine the free-space cyclotron frequency, $f_{c}$. Pertinent to the work described here is the relationship
\begin{equation}
f_{c} = f_{+} + f_{-} = \frac{qB}{2\pi m},
\end{equation}
where $q/m$ is the charge-to-mass ratio of the ion, and $B$ is the magnetic field strength. At LEBIT, the cyclotron frequency of ions in the Penning trap is determined using the time-of-flight ion cyclotron resonance (TOF-ICR) technique~\cite{Gra1980,kon1995}. Ions are parked on an initial magnetron orbit by a Lorentz steerer~\cite{Ringle2007} and are then driven with a quadrupolar radio frequency (rf) pulse at a frequency $f_{\textrm{rf}}$ close to the free-space cyclotron frequency of the ion. Depending on the frequency difference $\left|f_{\textrm{rf}} - f_{c}\right|$, magnetron motion can be partly or fully converted to cyclotron motion. The ions are then released from the trap and impinge on a micro-channel plate (MCP) detector that is used to determine the time-of-flight from the trap to the MCP, as well as the number of ions that were in the trap. An ion's time-of-flight to the MCP is minimized upon full conversion of magnetron to cyclotron motion. Hence, $f_{\textrm{rf}}$ is varied around $f_{c}$ and a characteristic TOF resonance curve can be mapped out. 

\begin{figure}[h]
\includegraphics[width=1\columnwidth]{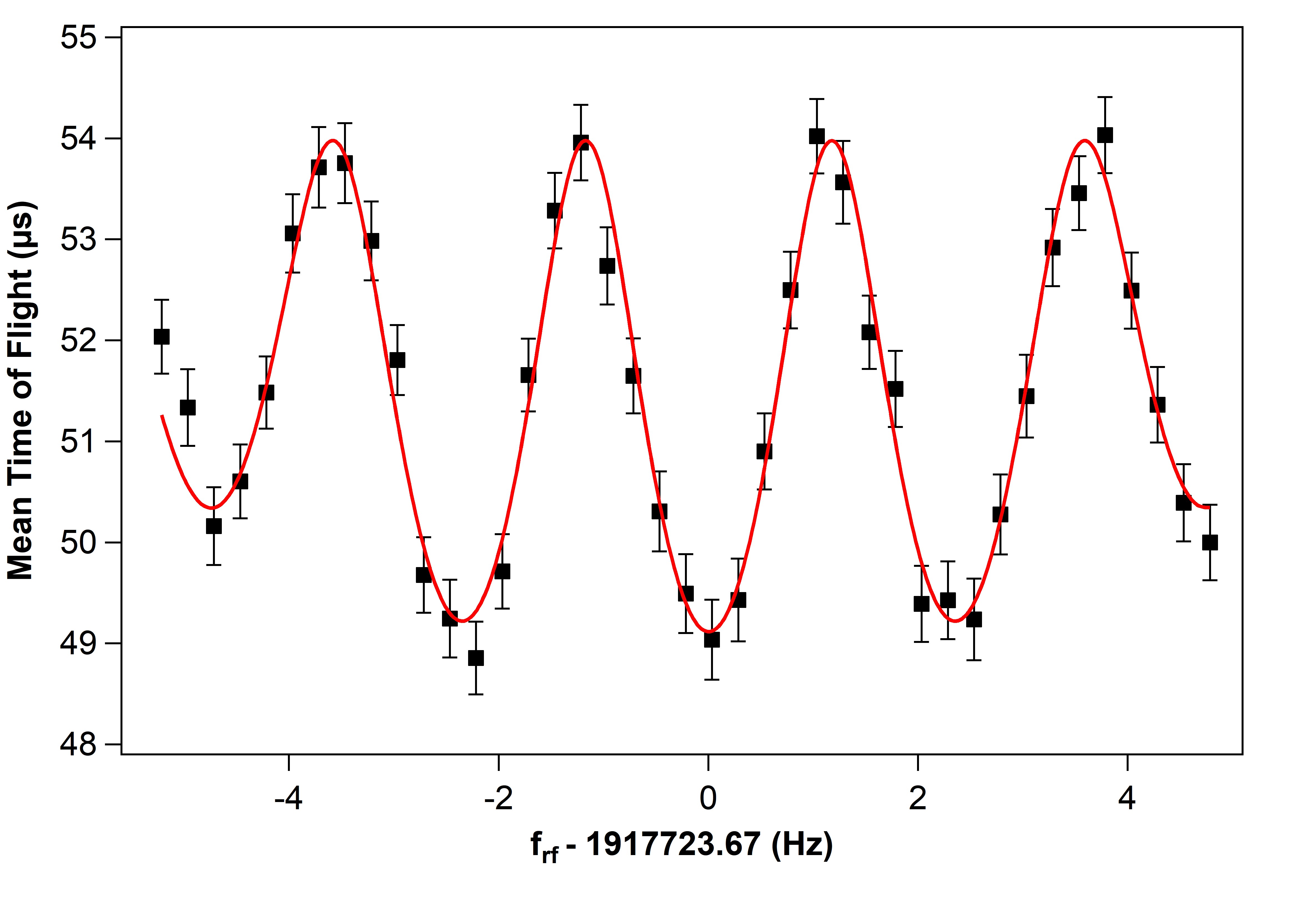}
\caption{(Color online) A  500 ms cyclotron frequency resonance curve for $^{75}$As$^{+}$. The solid red line is a fit to the theoretical line shape~\cite{Kretzschmar2007}.\label{fig:Resonance}} 
\end{figure}

In this work, we used the Ramsey TOF-ICR technique~\cite{Bollen1992,George2007}, in which two time separated $\pi$/2 pulses are used instead of a single $\pi$-pulse to couple the magnetron and cyclotron modes. This technique results in an interference-type pattern in the TOF resonance, and has been shown to enable a determination of the cyclotron frequency to at least 3 times higher precision in the same measurement time compared to the traditional TOF-ICR technique~\cite{George2007_PRL}. An example Ramsey TOF-ICR resonance for $^{75}$As$^{+}$ from this work is shown in Fig.~\ref{fig:Resonance}. 

A single resonance like the one shown in Fig.~\ref{fig:Resonance} took $\sim$30 mins, consisted of $\sim$2500 -- 5000 ions, and enabled a determination of the $^{75}$As$^{+}$ or $^{12}$C$_{6}^{+}$ frequency to $\sim$15 mHz. In order to determine the cyclotron frequency ratio of $^{12}$C$_{6}^{+}$ to $^{75}$As$^{+}$ and to account for temporal magnetic field drifts, we alternated between measurements with the two ions, performing 65 measurement pairs in total.

Deviations from an ideal Penning trap i.e. a magnetic field that is not perfectly uniform, and an electric field that is not purely quadrupolar, combined with the finite normal mode amplitudes of ions in the trap, result in shifts to the measured frequencies. The preparation of ions of different $m/q$ in the Penning trap can result in them having different normal mode amplitudes and hence different systematic frequency shifts. While these shifts largely cancel in the cyclotron frequency ratio, when comparing ions of different nominal $m/q$ a residual shift to the cyclotron frequency ratio can remain. In this work, our cyclotron frequency ratio involved $^{12}$C$_{6}^{+}$ and $^{75}$As$^{+}$, which differ by three atomic mass units. To evaluate the potential shift to the ratio, we took additional cyclotron frequency ratio data for $^{12}$C$_{6}^{+}$ vs $^{85,87}$Rb$^{+}$ and $^{12}$C$_{7}^{+}$. The uncertainty in the mass of the carbon cluster ions and the rubidium ions (the masses of which have been determined previously with the Florida State University Penning trap~\cite{Mount2010_Alkalis}) are negligible compared to the statistical uncertainties in our measurements. 

\section{Data and Analysis}
To evaluate the cyclotron frequency ratio from our alternating measurements of $^{12}$C$_{6}^{+}$ and $^{75}$As$^{+}$, pairs of $^{12}$C$_{6}^{+}$ cyclotron frequency measurements were first linearly interpolated to obtain the corresponding $f_c^{\textrm{int}}(^{12}$C$_{6}^{+})$ value at the time of the interleaved $f_c(^{75}$As$^{+})$ measurement. This procedure eliminated potential systematic shifts to the cyclotron frequency ratio due to linear magnetic field drifts. A series of cyclotron frequency ratios, corresponding to inverse mass ratios of the two ions were found from
\begin{equation}
R = \frac{f_c(^{75}\textrm{As}^{+})}{f_{c}^{\textrm{int}}(^{12}\textrm{C}_{6}^{+})} = \frac{m(^{12}\textrm{C}_{6}^{+})}{m(^{75}\textrm{As}^{+})}.
\end{equation}

\begin{figure}[h]
\includegraphics[width=1\columnwidth]{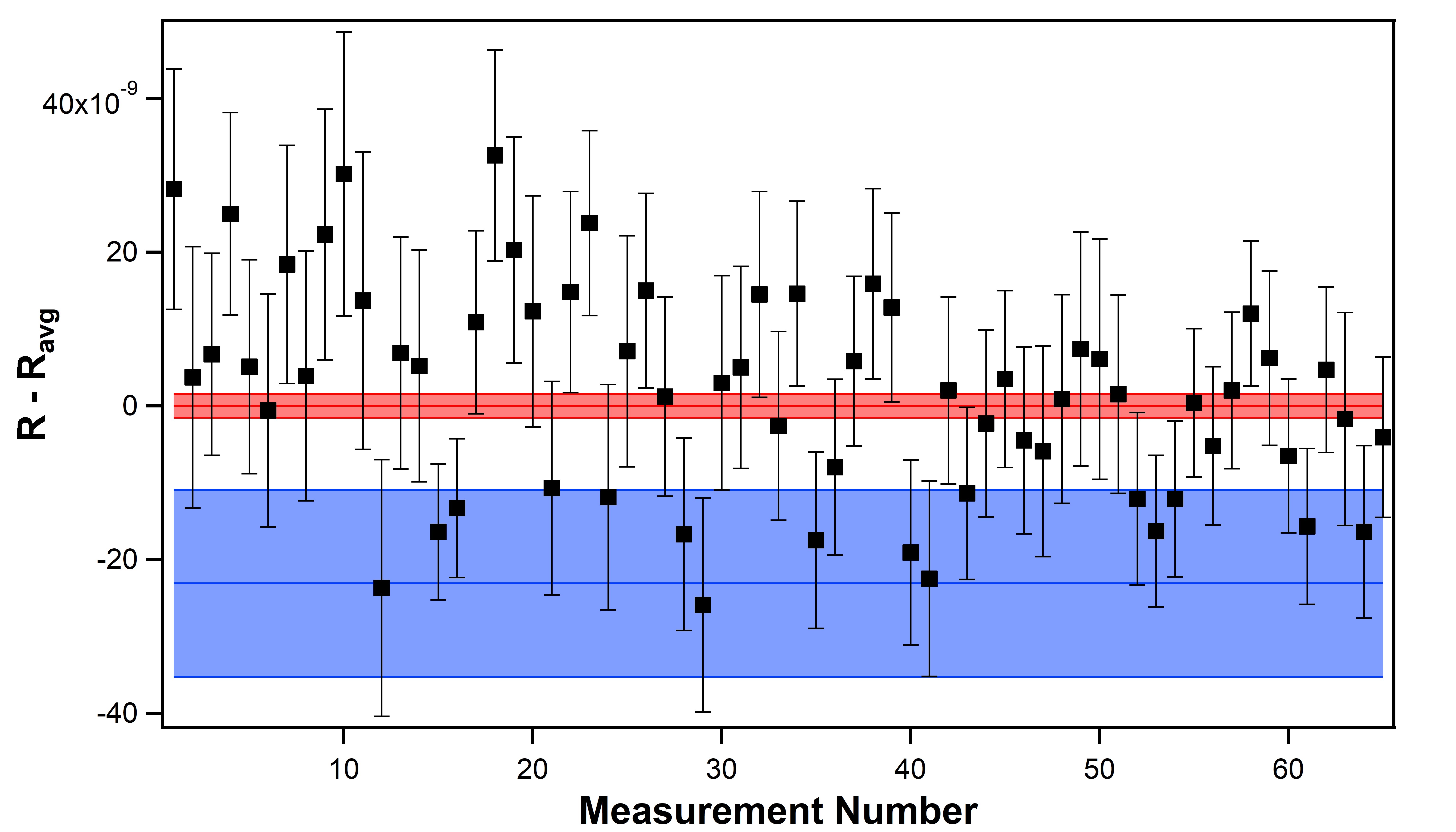}
\caption{(Color online) Individual cyclotron frequency ratio measurements of $f_{c}$($^{75}$As$^{+}$)/$f_{c}^{\textrm{int}}$($^{12}$C$_{6}^{+}$) (black squares). The red band shows the uncertainty in the average ratio, without systematic corrections and associated uncertainties applied. The blue line and band show the ratio obtained using the $^{75}$As mass from AME2020~\cite{Wang2021} and the associated uncertainty.\label{fig:Ratios}} 
\end{figure}

The individual measurements of $R$ were then used to find the weighted average, $\bar{R}$, the associated statistical uncertainty, and the Birge Ratio~\cite{Bir1932}, as shown in Fig.~\ref{fig:Ratios}. For the $^{12}$C$_{6}^{+}$ vs $^{75}$As$^{+}$ data set, consisting of 65 individual ratio measurements, the Birge Ratio was found to be 1.07(6) indicating that the statistical uncertainties in the individual measurements are a reasonable estimate of the statistical spread of the data. Nevertheless, we conservatively inflated the statistical uncertainty in the weighted average ratio by the Birge Ratio.

Our auxiliary measurements of $^{12}$C$_{6}^{+}$ vs $^{85,87}$Rb$^{+}$ and $^{12}$C$_{7}^{+}$ did show a systematic shift in the ratio when comparing ions of different nominal $m/q$ of --0.57(18) $\times$ 10$^{-9}$ per u/e. This result was used to apply a correction of 1.7(5) $\times$ 10$^{-9}$ to the $^{12}$C$_{6}^{+}$ vs $^{75}$As$^{+}$ ratio. We also performed a Z-class analysis to evaluate potential shifts to the cyclotron frequency from the Coulomb interaction when there is more than one ion in the trap~\cite{Kellerbauer2003}. This analysis showed a linear trend, but with a slope of 0.52(51) $\times$ 10$^{-9}$ per ion in the trap, consistent with no shift. In our data analysis, we limited the number of ions in the trap to between one and five. Assuming an average of three ions in the trap, a correction of --1.6(1.5) $\times$ 10$^{-9}$ should be applied to the data. We applied half this value as a correction with an uncertainty equal to the applied correction. 

In the course of the analysis we uncovered an additional systematic effect: when ions arrive at the MCP, they create a time-of-flight (TOF) distribution, the width of which depends on the energy difference between ions that are resonant with the rf quadrupole drive in the trap (i.e. magnetron motion is fully converted to cyclotron motion) and those that are not (i.e. no conversion). There is also some spread due to the variation in the initial conditions of ions in the trap. In addition, there can be background noise or ion events in the time window, and so a gate is placed around the TOF distribution of the ions of interest. For $^{75}$As$^{+}$ ions, a systematic variation in the cyclotron frequency was observed as a function of location of the high-end cut-off time of the gate i.e. on where the cut-off was made for ions with a longer TOF, which corresponds to non-resonant ions. This systematic effect was not observed for $^{12}$C$_{6}^{+}$ or $^{12}$C$_{7}^{+}$ ions produced with the LAS, or $^{85,87}$Rb$^{+}$ ions produced with a surface ionization source. Hence, we concluded that there could be contaminant ions in the beam arriving from the BMIS that were not identified, and which could cause systematic shifts to the $^{75}$As$^{+}$ ions' frequency via ion-ion interactions.

To account for this effect, we systematically varied the high-end cut-off time of the gate, and recorded the cyclotron frequency and $\chi^{2}$ of the fit averaged over all 65 measurements. We found that the $\chi^{2}$ varied quadratically around its minimum---when more ion events from the longer TOF tail of the distribution were included, the fit was statistically worse, and, likewise, when the cut was made further into the TOF distribution, more data were discarded and the fit was again worse. We used the gate-setting corresponding to the minimum $\chi^{2}$ (obtained from a fit of $\chi^{2}$ vs gate width) in our final analysis. The cyclotron frequency we obtained as a function of gate width also varied around a minimum value. The minimum $f_{c}$, however, did not occur at the minimum 
$\chi^{2}$. The minimum $f_{c}$ was 0.007 Hz lower, and the maximum $f_{c}$ 0.015 Hz higher than our nominal value at $\chi^{2}$-min. These correspond to shifts in the $^{75}$As$^{+}$/$^{12}$C$_{6}^{+}$ ratio of 3.7 and 7.7 $\times$ 10$^{-9}$, respectively. We used the average of these two values, 5.7 $\times$ 10$^{-9}$, to conservatively estimate the additional uncertainty to include in our final ratio to account for this systematic effect. This corresponds to an uncertainty in the mass of $^{75}$As of about 0.4 keV/$c^2$. A summary of our ratio measurement and the systematic corrections and uncertainties applied are shown in Table~\ref{table:ratio}.

\begin{table}[h]
\caption{\label{table:ratio} Error budget for $^{75}$As$^{+}$ vs $^{12}$C$_{6}^{+}$ cyclotron frequency ratio measurement.
$\sigma_{\mathrm{st}}$($\times$BR) is the statistical uncertainty (inflated by the  Birge Ratio, BR = 1.07). $\Delta R_\mathrm{ion}$, $\Delta R_\mathrm{\delta M}$, and $\Delta R_{\mathrm{gate}}$ are the systematic corrections (and corresponding uncertainties in parentheses) to account for frequency shifts due to there being more than one ion in the trap, the effect of comparing ions of dissimilar $m/q$, and the effect associated with how the gate on the TOF distribution was set (see text for details). $\bar{R}^{'}$ is the corrected ratio with statistical, systematic,and total uncertainties shown respectively in parentheses.}
\begin{ruledtabular}
\begin{tabular}{ccc}
\multirow{2}{*}{Source} & \multicolumn{2}{c}{Contribution to} \\
 & Ratio ($\times$ 10$^{-9}$) & $M$($^{75}$As) (keV/$c^{2}$) \\
\hline
$\sigma_\mathrm{st}$ & 1.5 & 0.11 \\
$\sigma_\mathrm{st}\times$BR & 1.6 & 0.12 \\
$\Delta R_\mathrm{ion}$ & --0.8(8) & 0.06(6) \\
$\Delta R_\mathrm{\delta M}$ & 1.7(5) & --0.12(4) \\
$\Delta R_\mathrm{gate}$ & 0.0(5.7) & 0.0(0.4) \\
\hline
Final ratio & \multicolumn{2}{c}{$\bar{R}^{'}$ = $0.961\ 004\ 378\ 4(16)(58)(60)$} \\
\end{tabular}
\end{ruledtabular}
\end{table}

After accounting for systematic effects, the corrected average ratio, $\bar{R}^{'}$, was used to determine the atomic mass of $^{75}$As in atomic mass units using the mass and binding energy of the carbon cluster reference molecule and accounting for the mass of the missing electrons, $m_e$:
\begin{equation}\label{MassEqn}
M(^{75}\textrm{As}) = \left[6M(^{12}\textrm{C}) - m_e + BE(^{12}\textrm{C}_{6})\right]\bar{R}^{'-1} + m_e.
\end{equation}
The first ionization energies of $^{75}$As and $^{12}$C$_{6}$ are both $\approx$10 eV, so the difference is $<$1 eV and is negligible. The $^{12}$C$_{6}$ molecular binding energy, on the other hand, is $-31.7$ eV for the most stable cyclic structure~\cite{Kos2008}, and therefore is included in Eqn.~(\ref{MassEqn}). The mass excess was then calculated from 
\begin{equation}
ME(^{75}\textrm{As}) = \left[M(^{75}\textrm{As})-75\right]\times F,
\end{equation}
where the conversion factor $F =
931\,494.102\,42(28)$ \textrm{(keV/$c^{2}$)/u} from Ref.~\cite{Tie2018} was used.

\section{Results and Discussion}

\begin{table}[b]
\caption{\label{table:mass} Mass excess for $^{75}$As obtained in the work along with the result from the AME2020~\cite{Wang2021} and the difference $\Delta$ME = ME$_{\rm{LEBIT}}$ -- ME$_{\rm{AME}}$}
\begin{ruledtabular}
\begin{tabular}{ccccc}
\multirow{2}{*}{Isotope} & \multirow{2}{*}{Ref.} & \multicolumn{1}{c}{This work} & \multicolumn{1}{c}{AME2020} & \multicolumn{1}{c}{$\Delta$ME}\\
 &  & \multicolumn{1}{c}{(keV/$c^2$)} & \multicolumn{1}{c}{(keV/$c^2$)} & \multicolumn{1}{c}{(keV/$c^2$)}\\
\hline
$^{75}$As & $^{12}$C$_{6}$ & $-73\ 035.98(43)$ & $-73\ 034.20(88)$ & --1.78(98)\\
\end{tabular}
\end{ruledtabular}
\end{table}

Our final mass excess value for $^{75}$As is given in Table \ref{table:mass}, and compared with the value from the AME2020~\cite{Wang2021}. Our result is 1.78(98) keV/$c^{2}$ lower than the AME2020 value and a factor of 2 more precise. This result indicates that $^{75}$As is more strongly bound than suggested by the AME2020 value, thus increasing the $Q_{gs}$ values of $^{75}$Se and $^{75}$Ge, which can be found via
\begin{equation}\label{75Se_Q}
Q^{\textrm{EC}}_{gs}\left(^{75}\textrm{Se}\right) = \left[ME\left(^{75}\textrm{Se}\right) - ME\left(^{75}\textrm{As}\right)\right]c^{2},
\end{equation}
and
\begin{equation}\label{75Ge_Q}
Q^{\beta}_{gs}\left(^{75}\textrm{Ge}\right) = \left[ME\left(^{75}\textrm{Ge}\right) - ME\left(^{75}\textrm{As}\right)\right]c^{2}.
\end{equation}
Our calculated $Q^{\textrm{EC}}_{gs}$ and $Q^{\beta}_{gs}$ values for $^{75}$Se and $^{75}$Ge, obtained using our new value for $M(^{75}$As) and the mass of the $^{75}$Se and $^{75}$Ge parent atoms from AME2020~\cite{Wang2021}, are given in Table III. 

The Q value for the decay to an excited state in the daughter nuclide with energy $E^{*}$ can then be found via
\begin{equation}\label{Q_UL}
  Q_{es}^{\beta/\textrm{EC}} = Q_{gs}^{\beta/\textrm{EC}} - E^{*}.    
\end{equation}

For the case of $^{75}$Se EC decay, two potential excited states in $^{75}$As with energies close to $Q^{\textrm{EC}}_{gs}(^{75}$Se) were identified: the 860.0(4) keV 1/2$^{+}$ level, and the 865.4(5) keV 3/2$^{+}$ or 5/2$^{+}$ level. For $^{75}$Ge $\beta$-decay, the 1172.0(6) keV $J^{-}$ state with $J$ = 1/2, 3/2, 5/2 or 7/2 was identified as a potential daughter state with a low energy decay Q value. Our $Q_{gs}$ values are compared with the relevant $E^{*}$ values in Table III. Our results indicate that all three decays listed are energetically allowed at least at the 1.5$\sigma$ level. The $^{75}$Ge $\beta$-decay to the 1172 keV state in $^{75}$As is ruled out as not being ultra-low (i.e. $Q_{es}^{\beta}$ $>$ 1 keV) at the 9$\sigma$ level. Likewise, the EC decay of $^{75}$Se to the 860 keV level in $^{75}$As is ruled out as an ultra-low decay branch, at the 8$\sigma$ level. However, the EC decay of $^{75}$Se to the 865 keV level in $^{75}$As has $Q_{es}^{\textrm{EC}} \approx$ 1 keV, making it a potential candidate for an ultra-low Q value decay. The uncertainty in the $Q_{es}^{\textrm{EC}}$ value listed in Table III is predominantly (60 \% contribution) due to the 0.5 keV uncertainty in the 865 keV level, with the remaining 40 \% contribution due to the 0.4 keV uncertainty in our $M(^{75}$As) value. Hence, a more precise determination of both the $Q^{\textrm{EC}}_{gs}(^{75}$Se) value and the $E^{*}$(865.4 keV) level energy are required to further evaluate this potential ultra-low Q value EC decay. 

In fact, a recent measurement reported on the arXiv by the JYFLTRAP group does provide a more precise mass for $^{75}$As~\cite{Ram2022}. Our result for $ME$($^{75}$As) agrees with the JYLFTRAP result, $ME$($^{75}$As) = --73035.521(42) keV/$c^{2}$, which is a factor of ten more precise, at the 1$\sigma$ level. The JYFLTRAP results also indicates that $^{75}$Se EC decay to the 865.4 keV level in $^{75}$As is energetically allowed with a potentially ultra-low Q value, $Q_{es} \lesssim$ 1 keV, but that a more precise determination of the 865.4 keV level is necessary.

\begin{table}[t]
 \caption{\label{table:Q} Q values based on the mass measurement listed in Table~\ref{table:mass} and Eqns.~(\ref{75Se_Q}) and (\ref{75Ge_Q}). The column $E^*$ gives the energy of the excited state of the daughter nucleus that the potential ultra-low Q value decay would go to. The result for the Q value of the decay branch to the excited state, $Q_{es}$, is calculated from Eqn.~(\ref{Q_UL}).}
\begin{ruledtabular}
\begin{tabular}{ccccc}
\multirow{2}{*}{Parent} & \multirow{2}{*}{Decay} & \multicolumn{1}{c}{$Q_{gs}$} & \multicolumn{1}{c}{$E$*} & \multicolumn{1}{c}{$Q_{es}$} \\
 & & \multicolumn{1}{c}{(keV)} & \multicolumn{1}{c}{(keV)} & \multicolumn{1}{c}{(keV)} \\
\hline
$^{75}$Se & EC & 866.50(44) & 865.4(5) & 1.10(67) \\
$^{75}$Se & EC & 866.50(44) & 860.4(4) & 6.10(60) \\
$^{75}$Ge & $\beta$ & 1179.01(44) & 1172.0(6) & 7.01(74)\\
\end{tabular}
\end{ruledtabular}
\end{table}

\section{Conclusion}

Using Penning trap mass spectrometry, we have performed a precise measurement of the mass of $^{75}$As. Our result reduces the uncertainty in the mass by about a factor of two compared to the value adopted in the most recent atomic mass evaluation, AME2020~\cite{Wang2021} and shows a shift of --1.78 keV/$c^{2}$. Using this new mass and the more precisely known masses of $^{75}$Se and $^{75}$Ge from AME2020, we were able to determine new ground-state to ground-state Q values for the EC decay of $^{75}$Se and the $\beta$-decay of $^{75}$Ge. These Q values were combined with excited state energies in $^{75}$As to determine the $Q_{es}$ values of potential ultra-low energy decays of $^{75}$Se and $^{75}$Ge. Our results indicate that the $\beta$-decay of $^{75}$Ge to the 1172.0(6) keV, negative parity state with angular momentum between 1/2 and 7/2, is energetically allowed, but with a Q value of 7.0(7) keV, ruling it out as a potential ultra-low Q value decay. Likewise, the EC decays of $^{75}$Se to the 865.4(5) keV 3/2$^{-}$ or 5/2$^{-}$ level, and to the 860.0(4) keV 1/2$^{+}$ level are also energetically allowed with Q values of 1.1(7) keV and 6.1(6) keV, respectively. Our result is in agreement with a recently reported measurement by the JYFLTRAP group. Both results indicate that the EC decay of $^{75}$Se to the 865 keV level in $^{75}$As with a $Q_{es}$ $\lesssim$ 1 keV  is of interest as a potential ultra-low Q value decay candidate and that, in order to further evaluate this potential decay branch, a more precise determination of the energy of the 865 keV level is required. 

\section*{Acknowledgments}

This research was supported by Central Michigan University, Michigan State University and the Facility for Rare Isotope Beams, and the National Science Foundation under Contracts No. PHY-1565546, PHY-2111185, and No. PHY-2111302.

\bibliography{75As_paper.bib} 

\end{document}